\documentclass[12pt]{iopart}

\usepackage{epsfig, iopams, graphicx}

\usepackage{cancel}
\usepackage{import}
\usepackage{float}
\usepackage{url}
\usepackage{color}

\newcommand\bb[1] {   \mbox{\boldmath{$#1$}}  }

\newcommand\del{\bb{\nabla}}

\def\gtsima{$\; \buildrel > \over \sim \;$}
\def\gtsim{\lower.5ex\hbox{\gtsima}}

\newcommand\bcdot{\bb{\cdot}}

\newcommand\dd{\partial}

\newcommand\hmn { {\bar h_{\mu\nu}} }
\newcommand\Tmn { {T_{\mu\nu}} }
\newcommand\Smn { {S_{\mu\nu}} }

\newcommand\SSS { {\cal S} }
\newcommand\kk{ \kappa }
\newcommand\eps{\epsilon}
\newcommand\beq{ \begin{equation} }
\newcommand\eeq{ \end{equation} }

\begin{document}
\title[Gravitational wave stress tensor from the linearized field equations]{\bf \Large Simplified derivation of the gravitational wave stress tensor from the linearized Einstein field equations}

\author{ Steven A. Balbus}

\address {Department of Physics, Astrophysics, University of Oxford, 
Denys Wilkinson Building, Keble Road, Oxford OX13RH}
  \ead{steven.balbus@physics.ox.ac.uk}



\begin{abstract}
A conserved stress energy tensor for weak field gravitational waves propagating in vacuum is derived directly from the linearized wave equation alone, for an arbitrary gauge using standard general relativity.   In any harmonic gauge, the form of the tensor leads directly to the classical expression for the outgoing wave energy.   The method described here, however, is a much simpler, shorter, and more physically motivated approach than is the customary procedure, which involves a lengthy and cumbersome second-order (in wave-amplitude) calculation starting with the Einstein tensor.   Our method has the added advantage of exhibiting the direct coupling between the outgoing energy flux in gravitational waves and the work done by the gravitational field on the sources.   For nonharmonic gauges, the directly derived wave
stress tensor has an apparent index asymmetry.   This coordinate artefact may be removed, and 
the symmetrized (still gauge-invariant) tensor then takes on its widely-used form.   Angular momentum conservation follows immediately.   For any harmonic gauge, however, the stress tensor is manifestly symmetric from the start and its derivation depends, in its entirety, on the structure of the linearized wave equation. 
\end{abstract}


\medskip

\noindent {\small Publication Reference:   2016, PNAS,  113, 42, 11662-11666.   doi: 10.1073/pnas.$\,$1614681113}

\maketitle
\section*{Significance}

Gravitational radiation provides a probe of unprecedented power with which to elucidate important astrophysical
processes otherwise completely dark (e.g. black hole mergers) or impenetrable
(e.g. supernovae and early universe dynamics).    Historically, the gap between propagating fluctuations in the spacetime metric
and classical dynamical concepts such as energy and angular momentum conservation
has bedeviled this subject.    By now there is a vast literature on this topic, and there are many
powerful methods available.  Because of their mathematical sophistication, however,
they are not generally used in introductory texts, which are forced instead to a follow a much more cumbersome path.   We present here a derivation of the most widely use form of the stress energy tensor of gravitational radiation, using elementary methods only.   

\section{Introduction} 

The recent detection of gravitational radiation \cite{abb} has greatly heightened interest in this subject.  Deriving an expression for the correct form of the energy flux carried off in the form of gravitational waves is a famously difficult undertaking at both the
conceptual and technical levels.   The heart of the difficulty is that the stress energy of the gravitational field is neither a unique nor a localizable quantity, because local coordinates can be found for which the field can be made to vanish by the equivalence principle.   It is not a source of spacetime curvature; it is part of the curvature itself, which manifests globally.   Indeed, for many years, debate abounded as to whether
there was any true energy propagated by gravitational radiation.   We now know of course that there is, but the hunt for a suitable stress energy is a burdensome demand for those approaching the subject either as nonspecialists or as newcomers.   The currently generally adopted textbook approach\cite{mag} is to first write the metric tensor $g_{\mu\nu}$ as the following sum\cite{note}: 
\beq
g_{\mu\nu} = \eta_{\mu\nu} + h_{\mu\nu},
\eeq
where $\eta_{\mu\nu}$ is the usual Minkowski metric and $h_{\mu\nu}$ the departure therefrom, and then to treat the latter as a small quantity.    We work throughout in quasi-Cartesian coordinates that differ only infinitesimally in linear order from strictly Cartesian
coordinates, so that $\eta_{\mu\nu}$ is a constant tensor (see endnote).   The Einstein tensor,
\beq
G_{\mu\nu} \equiv R_{\mu\nu} - {g_{\mu\nu}R\over 2},
\eeq
where $R_{\mu\nu}$ is the Ricci tensor and $R$ its $R^\mu_{\ \mu}$ trace, is then expanded in powers of the amplitudes of $h_{\mu\nu}$ in its various forms.   With the material stress energy tensor denoted by $T_{\mu\nu}$, the Newtonian gravitational constant by $G$, and the speed of light set to unity, the Einstein field equation is the following:
\beq
G_{\mu\nu} = - 8\pi GT_{\mu\nu},
\eeq
which upon expansion in $h_{\mu\nu}$ may be rewritten as follows:
\beq
G^{(1)}_{\mu\nu} = - 8\pi G (T_{\mu\nu} +t_{\mu\nu}).
\eeq
Here $G^{(1)}_{\mu\nu}$ consists of the terms in $G_{\mu\nu}$ that are linear in $h_{\mu\nu}$, and 
\beq
t_{\mu\nu} = {1\over 8\pi G} (G^{(2)}_{\mu\nu} +...),
\eeq
where $G^{(2)}_{\mu\nu}$ represents the Einstein tensor terms quadratic in $h_{\mu\nu}$, and so forth.     Following a
standard practice, we refer to $t_{\mu\nu}$ as a ``pseudotensor,''  because it is Lorentz covariant but not a true tensor, unlike $T_{\mu\nu}$, under full coordinate transformations.  To leading nonvanishing order,  the pseudotensor $t_{\mu\nu}$ is then interpreted as the stress energy of the gravitational radiation itself.  
The sum $T_{\mu\nu}+t_{\mu\nu}$ is often referred to as the energy-momentum pseudotensor;
a yet more general version of the pseudotensor is presented in the text of Landau \& Lifschitz\cite{ll}.    There are by now many routes that lead to a suitable definition of an appropriate stress tensor
for gravitational radiation, without the use of the pseudotensor formalism.  We make no pretense of doing anywhere near full justice to this elegant and sophisticated literature here; this is not the intent of this article.   Our purpose, rather, is to show how to obtain a widely-used form of the stress-energy tensor for gravitational radiation, making use only
of elementary methods and conserved fluxes emerging from linear wave theory.   

The calculation of the energy from the pseudotensor is sufficiently 
cumbersome that it is rarely done explicitly in textbooks (merely summarized), although the final answer is not unduly involved.   In an arbitrary gauge and a background Minkowski spacetime\cite{mis},
$$
t_{\mu\nu} = {1\over 32\pi G}\Bigg[ \left\langle 
{\dd \bar h_{\kappa\lambda}\over  \dd x^\mu}   {\dd{\bar h}^{\kappa\lambda}\over \dd x^\nu}   \right\rangle- \left\langle {\dd {\bar h}^{\lambda\kappa}\over  \dd x^\lambda}{\dd {\bar h}_{\kappa\mu}\over \dd x^\nu}\right\rangle- \left\langle {\dd {\bar h}^{\lambda\kappa}\over  \dd x^\lambda}{\dd {\bar h}_{\kappa\nu}\over \dd x^\mu}\right\rangle  \ \ \ \ \ \ \ \ \ \ \ \ \ \ \  \ \ \ \ \ \  \ \ \ \ \ \ 
$$
\beq\label{tt}
 \qquad\qquad\quad\qquad\qquad\qquad\qquad\qquad\qquad\qquad -{1\over 2}\left\langle{\dd \bar h\over \dd x^\mu}
{\dd \bar h\over \dd x^\nu}\right\rangle\Bigg],
\eeq
where
$$
\hmn = h_{\mu\nu} - {\eta_{\mu\nu}\over 2}h, \quad h\equiv h^\mu_\mu, \quad \bar h \equiv \bar h^\mu_\mu=-h.
$$
As explained in standard texts, the use of equation (\ref{tt}) as a stress tensor makes sense only if an average over many wavelengths is performed, so that oscillatory cross products do not contribute.   This averaging is indicated by the angle bracket $\langle\rangle$ notation.    Moreover, although the expression (\ref{tt}) is gauge-invariant,
in solving explicitly for $h_{\mu\nu}$, a choice of gauge must be made.  The ``harmonic gauge'' is a convenient
choice for the study gravitational waves, as it greatly simplifies the mathematics.  If $h_{\mu\nu}$  depends on its coordinates as a plane wave of the form $\exp(ik_\mu x^\mu)$,  an harmonic gauge is actually required if $k_\mu$ is a null vector,
$k_\mu k^\mu=0$.   All physical, curvature-inducing radiation (as opposed to oscillating coordinate transformations) has this property\cite{mis}.  The harmonic gauge is defined by the condition 
\beq
{\dd\hmn\over \dd x_\mu} = 0 \quad {\rm (harmonic\ gauge\ condition).}
\eeq
That it is always possible to find such a gauge is well-known\cite{mis}; the proof is similar to
that of being able to choose the Lorenz gauge condition in electrodynamics.   
In the ``transverse traceless'' (TT) gauge, there is the additional constraint $h=0$, which leads to the simple result
\beq\label{ttg}
t_{\mu\nu}={1\over 32\pi G}\left\langle 
{\dd \bar h_{\kappa\lambda}\over  \dd x^\mu}   {\dd{\bar h}^{\kappa\lambda}\over \dd x^\nu}   \right\rangle\quad{\rm (TT\ gauge).}
\eeq
For linear gravitational plane waves propagating in vacuum (though not more generally\cite{mis}), a transformation to the TT gauge can always be found without departing from the harmonic constraint;  there is also a precise electrodynamic counterpart.   

By way of contrast, in classical wave problems, finding a conserved wave energy flux is much more straightforward. 
Consider the simplest example of a wave equation for a quantity $f$,
\beq
{\dd ^2 f\over \dd t^2 } - {\dd^2  f\over \dd x^2} = 0.
\eeq
Start by looking for a conserved flux.   If we multiply by ${\dd f/ \dd t} \equiv \dot f$, integrate the second term $-\dot f \dd^2 f/\dd x^2$ by parts and regroup, this leads to the following:
\beq
{\dd\ \over \dd t} \left [ {{\dot f}^2\over 2} +{(f')^2\over 2} \right] -{\dd\ \over \dd x} \left(\dot f f'\right)=0,
\eeq
where $f'\equiv {\dd f/\dd x}$.  This readily lends itself to the interpretation of an energy density $[\dot f^2 +f'^2]/2$ and an
energy flux $-\dot f f'$, though with an uncertain overall normalization factor that must be determined by such considerations
as the work done on the wave sources.    Note in particular that the energy flux is second order in the $f$-amplitude, even though it is ultimately
determined by a linear-in-$f$ wave equation.  

A linear scalar wave equation is yet more revealing, and only slightly more complicated.  With $\Phi$ the effective potential and $\rho$ the source density, consider the wave equation of scalar gravity,
\beq
\Box\Phi \equiv -{\dd ^2 \Phi \over \dd t^2 } +\nabla^2\Phi   = 4\pi G\rho.
\eeq
(Here $\Box$ and $\nabla^2$ are the usual d'Alembertian and Laplacian operators respectively.)
Then, if we multiply by $(1/4\pi G){\dd_t \Phi}$, integrate $(\dd_t\Phi) \nabla^2\Phi$ by parts and regroup, this leads to the following:
\beq
-{1\over 8\pi G} {\dd\ \over \dd t} \left [ \left(\dd\Phi\over \dd t\right)^2  +\left|\del\Phi\right|^2 \right] +\del\bcdot\left({1\over 4\pi G}
{\dd\Phi\over \dd t} \del\Phi\right)=\rho{\dd\Phi\over \dd t}.
\eeq
However,
\begin{eqnarray}
\rho{\dd\Phi\over \dd t}&=& {\dd(\rho\Phi)\over \dd t}-\Phi{\dd\rho\over \dd t}\\
  & = &   {\dd(\rho\Phi) \over \dd t} +\Phi\del\bcdot(\rho\bb{v})=
{\dd(\rho\Phi) \over \dd t}  +\del\bcdot(\rho\bb{v}\Phi) -\rho\bb{v}\bcdot\del\Phi,
\end{eqnarray}
where $\bb{v}$ is the velocity and the usual mass conservation equation has been used in the second equality.  A simple rearrangement
then leads to the following:
\beq\label{scale}
 {\dd{\cal E} \over \dd t} +\del\bcdot{\bb{F}} =\rho\bb{v}\bcdot\del\Phi,
\eeq
where
\beq
{\cal E} =  \rho\Phi +{1\over 8\pi G}\left[\left(\dd\Phi\over \dd t\right)^2  +\left|\del\Phi\right|^2 \right], \ \ 
\bb{ F}=\rho\bb{v}\Phi -{1\over 4\pi G}{\dd\Phi\over \dd t} \del\Phi
\eeq
The right side of (\ref{scale}) is minus the volumetric rate at which work is being done on the sources.  For the usual case of compact sources, the left side may then be interpreted as a far-field wave energy density of $[(\dd_t\Phi)^2+|\del\Phi|^2]/8\pi G$ and a wave energy flux of
$-(\dd_t\Phi) \del\Phi/4\pi G$.   The question we raise here is whether an analogous ``direct method'' might be used to shed
some light on the origin of equation (\ref {tt}), including, very importantly, a means of extracting the overall normalization factor.   

There does indeed seem to be such a formulation, which we now discuss.   

\section{Analysis}

\subsection {Conserved densities and fluxes}

Begin with the standard, gauge-invariant general weak field linearized wave equation\cite{mis,mag}:
\beq\label{1}
\Box \hmn- {\dd^2 {\bar h}^\lambda_\mu\over \dd x^\nu \dd x^\lambda} - {\dd^2 {\bar h}^\lambda_\nu\over \dd x^\mu \dd x^\lambda} +\eta_{\mu\nu} {\dd^2{\bar h}^{\lambda\rho}\over \dd x^\lambda \dd x^\rho }=-\kk \Tmn,
\eeq
where $\kk=16\pi G$.  We restrict our attention throughout this work to the case of a small metric disturbance $h_{\mu\nu}$ on a background Minkowski spacetime.  The material stress tensor $T_{\mu\nu}$ is treated as completely Newtonian.

Next, establish an identity by contracting equation (\ref{1}) on $\mu\nu$:
$$
\Box\bar h + 2 {\dd^2{\bar h}^{\lambda\rho}\over \dd x^\lambda \dd x^\rho }=-\kk T^\mu_{\ \mu}\equiv -\kk T.
$$
Hence:
\beq
{\dd^2{\bar h}^{\lambda\rho}\over \dd x^\lambda \dd x^\rho }= -{1\over 2} \Box\bar h -{\kk T\over 2},
\eeq
and we rewrite our equation as 
\beq\label{eq}
\Box \hmn- {\dd^2 {\bar h}^\lambda_\mu\over \dd x^\nu \dd x^\lambda} - {\dd^2 {\bar h}^\lambda_\nu\over \dd x^\mu \dd x^\lambda} -{\eta_{\mu\nu}\over 2}\Box\bar h =-\kk \Smn,
\eeq
where the source function $\Smn$ is
\beq
\Smn = \Tmn -{{\eta}_{\mu\nu}T\over 2}.
\eeq

We seek an energy-like conservation equation from the wave equation in the form displayed in equation (\ref{eq}).   Towards that end, multiply by $\dd_\sigma {\bar h}^{\mu\nu}$, summing over $\mu\nu$ as usual and leaving $\sigma$ free.    The first term on the left side of (\ref{eq}) is then
$$
{\dd^2\hmn\over \dd x^\rho\dd x_\rho} {\dd {\bar h}^{\mu\nu}\over \dd x^\sigma} = {\dd\over \dd x_\rho}\left( 
{\dd\hmn\over  \dd x^\rho}   {\dd{\bar h}^{\mu\nu}\over \dd x^\sigma}   \right) - {\dd\hmn\over  \dd x^\rho}{\dd^2{\bar h}^{\mu\nu}
\over \dd x_\rho\dd x^\sigma},
$$
$$
={\dd\ \over \dd x_\rho}\left( 
{\dd\hmn\over  \dd x^\rho}   {\dd{\bar h}^{\mu\nu}\over \dd x^\sigma}   \right) - {\dd\hmn\over  \dd x^\rho}{\dd\ \over \dd x^\sigma}
{{\dd \bar h}^{\mu\nu}\over \dd x_\rho},
$$
\beq 
={\dd\ \over \dd x_\rho}\left( 
{\dd\hmn\over  \dd x^\rho}  {\dd{\bar h}^{\mu\nu}\over \dd x^\sigma}   \right) - {1\over 2} {\dd\ \over \dd x^\sigma}\left(
{\dd\hmn\over  \dd x^\rho}{{\dd \bar h}^{\mu\nu}\over \dd x_\rho}\right).
\eeq
The second term on the left is handled similarly.  Juggling indices,
\beq
-{\dd^2 {\bar h}^\lambda_\mu\over \dd x^\nu \dd x^\lambda}{\dd {\bar h}^{\mu\nu}\over \dd x^\sigma}=
-{\dd^2 {\bar h}^{\lambda\mu}\over \dd x_\nu \dd x^\lambda}{\dd {\bar h}_{\mu\nu}\over \dd x^\sigma},
\eeq
leads to the following:
\beq
-{\dd^2 {\bar h}^\lambda_\mu\over \dd x^\nu \dd x^\lambda}{\dd {\bar h}^{\mu\nu}\over \dd x^\sigma}=
-{\dd\ \over \dd x_\nu}\left( {\dd {\bar h}^{\lambda\mu}\over  \dd x^\lambda}{\dd {\bar h}_{\mu\nu}\over \dd x^\sigma}\right)
+ {\dd {\bar h}^{\lambda\mu}\over  \dd x^\lambda} {\dd^2 {\bar h}_{\mu\nu}\over\dd x^\sigma \dd x_\nu},
\eeq
or equivalently,
\beq
-{\dd^2 {\bar h}^\lambda_\mu\over \dd x^\nu \dd x^\lambda}{\dd {\bar h}^{\mu\nu}\over \dd x^\sigma}=-{\dd\ \over \dd x_\rho}\left( {\dd {\bar h}^{\lambda\mu}\over  \dd x^\lambda}{\dd {\bar h}_{\mu\rho}\over \dd x^\sigma}\right)+{1\over 2}{\dd \over \dd x^\sigma}
\left( {\dd {\bar h}^{\lambda\mu}\over  \dd x^\lambda} {\dd\hmn\over \dd x_\nu}\right).
\eeq
The third term is identical to the second upon summmation over $\mu$ and $\nu$.
The fourth and final term of the left side of equation (\ref{eq}) is 
\beq
- {1\over 2} {\dd^2\bar h\over \dd x^\rho\dd x_\rho} {\dd {\bar h}\over \dd x^\sigma}= - {1\over 2}{\dd\ \over \dd x_\rho}\left({\dd \bar h\over \dd x^\rho}
{\dd \bar h\over \dd x^\sigma}\right)+ {1\over 4} {\dd\ \over \dd x^\sigma}\left( {\dd\bar h\over \dd x^\rho}{\dd \bar h\over \dd x_\rho}
\right).
\eeq
Thus, after dividing by $2\kappa$, equation (\ref{eq}) takes the form of 
\beq\label{wee}
{\dd \SSS\over \dd x^\sigma} + {\dd {\cal T}_{\rho\sigma}\over \dd x_\rho} = -{1\over 2} \Smn  {\dd{\bar h}^{\mu\nu}\over \dd x^\sigma},
\eeq
where $\SSS$ is a scalar density:
\beq\label{S}
\SSS = -{1\over 4\kk}\left( 
{\dd\hmn\over  \dd x^\rho}{{\dd \bar h}^{\mu\nu}\over \dd x_\rho}\right)+ {1\over 2\kk}\left( {\dd {\bar h}^{\lambda\mu}\over  \dd x^\lambda} {\dd\hmn\over \dd x_\nu}\right)+{1\over 8\kk}\left( {\dd\bar h\over \dd x^\rho}{\dd \bar h\over \dd x_\rho}
\right),
\eeq
and ${\cal T}_{\rho\sigma}$ is a flux tensor:
\beq\label{T}
{\cal T}_{\rho\sigma} = {1\over 2\kk}\left( 
{\dd\hmn\over  \dd x^\rho}   {\dd{\bar h}^{\mu\nu}\over \dd x^\sigma}   \right)-{1\over \kk} \left( {\dd {\bar h}^{\lambda\mu}\over  \dd x^\lambda}{\dd {\bar h}_{\mu\rho}\over \dd x^\sigma}\right)- {1\over 4\kk}\left({\dd \bar h\over \dd x^\rho}
{\dd \bar h\over \dd x^\sigma}\right).
\eeq
Due to its second term, ${\cal T}_{\rho\sigma}$ is not symmetric in its indices.  Index asymmetry also arises in the development of
the stress tensor of electromagnetic theory, and there are methods to correct this deficiency \cite{jac}.   Similar techniques may be brought to bear on the current problem, as we discuss below.   For the moment, we may note that in an harmonic gauge $\dd_\mu \bar h^{\mu\nu}=0$ (not necessarily traceless), the
asymmetry vanishes and the tensor becomes manifestly symmetric in $\rho\sigma$.   
Notice that the wave stress tensor (\ref{tt}) is simply a symmetrized version of (\ref{T}).  

Rather than work with ${\cal S}$ and ${\cal T}_{\rho\sigma}$ each on its own, it is more natural to form the composite tensor ${\cal U}_{\rho\sigma}$,
\beq\label{u}
{\cal U}_{\rho\sigma} \equiv  {\cal T}_{\rho\sigma} + \eta_{\rho\sigma}{\cal S}.
\eeq
The left side of equation (\ref{wee}) may then be written more compactly as a 4-divergence:
\beq\label{we2}
{\dd {\cal U}_{\rho\sigma} \over \dd x_\rho} =  -{1\over 2} \Smn  {\dd{\bar h}^{\mu\nu}\over \dd x^\sigma}.
\eeq
It should be noted that the content of equation (\ref{we2})
is exactly the same as that of the wave equation (\ref{1}): no more, no less.  At this stage, note that we have not done
any spatial averaging.   In the TT gauge equation (\ref{u}) leads directly to:
\beq
{\cal U}_{00} = {1\over 4\kk}\left(
{\dd\hmn\over  \dd x^i}{{\dd \bar h}^{\mu\nu}\over \dd x^i}+
  {\dd\hmn\over  \dd t}   {\dd{\bar h}^{\mu\nu}\over \dd t}   \right),
\eeq
\beq
{\cal U}_{0 i } =  {1\over 2\kk}\left(
{\dd\hmn\over  \dd t}   {\dd{\bar h}^{\mu\nu}\over \dd x^i}   \right),
\eeq
\beq
{\cal U}_{ij} = {1\over 2\kk}\left( 
{\dd\hmn\over  \dd x^i}   {\dd{\bar h}^{\mu\nu}\over \dd x^j}  -
 {{\delta}_{ij} \over 2}
{\dd\hmn\over  \dd x^\rho}{{\dd \bar h}^{\mu\nu}\over \dd x_\rho}\right).
\eeq
By these canonical forms, the component ${\cal U}_{00}$ is readily interpreted as a wave energy density, ${\cal U}_{0i}$ as a wave energy flux, and ${\cal U}_{ij}$ as a wave momentum stress.  But in fact, the combination $({\dd^\rho{\bar h}_{\mu\nu})(\dd_\rho \bar h}^{\mu\nu})$ (and $[{\dd^\rho{\bar h}][\dd_\rho \bar h}]$ in a more general harmonic gauge) must vanish when averaged over many wavelengths, since the fourier wave vector
components satisfy the null constraint $k^\rho k_\rho = 0$.    In the end, there emerges the very simple results
\begin{eqnarray}\label{summ}
{\cal U}_{\rho\sigma}   =  t_{\rho\sigma} &  =& {1\over 2 \kk} \left\langle 
{\dd\hmn\over  \dd x^\rho}   {\dd{\bar h}^{\mu\nu}\over \dd x^\sigma}-{1\over 2} {\dd \bar h\over  \dd x^\rho}   {\dd{\bar h}\over \dd x^\sigma} \right\rangle \quad\, {\rm (harmonic\ gauge),} \nonumber\\
 & = & {1\over 2 \kk} \left\langle 
{\dd\hmn\over  \dd x^\rho}   {\dd{\bar h}^{\mu\nu}\over \dd x^\sigma} \right\rangle \qquad\qquad\qquad\quad\!  {\rm (TT\ gauge).}
\end{eqnarray}

\subsection {Direct energy loss}

In going from the wave equation (\ref{eq}) to an energy equation (\ref{we2}) we divided by $2\kk$.   How do we know that this particular normalisation is the proper one for producing a true energy flux?   It is the right side of equation (\ref{wee}) that tells this story.
This is
\begin{eqnarray}
-{1\over 2}S_{\mu\nu} {\dd{\bar h}^{\mu\nu}\over \dd x^\sigma} & = & -{1\over 2}\left(T_{\mu\nu} - {\eta_{\mu\nu}\over 2}T\right)
\left( {\dd h^{\mu\nu}\over \dd x^\sigma}-{\eta^{\mu\nu}\over 2}{\dd  h\over \dd x^\sigma}\right)\nonumber\\
&=& -{1\over 2}T_{\mu\nu} {\dd  h^{\mu\nu}\over \dd x^\sigma}.
\end{eqnarray}
We now set $\sigma = 0$, picking out the time component, and work in the Newtonian limit $h^{00} \simeq -2\Phi$, where $\Phi$ is 
the gravitational potential.   We are then dominated by the $00$ components of $h^{\mu\nu}$ and $T_{\mu\nu}$.   Using the right arrow $\rightarrow$ to mean integrate by parts and ignore the pure derivatives (as inconsequential for wave losses), and
recalling the mass-energy conservation relation $\dd_\mu T^{ 0\mu}=0$, we perform the following manipulations:
\begin{eqnarray}\label{32}
-{1\over 2}T_{00} {\dd  h^{00}\over \dd x^0}\rightarrow {1\over 2}{\dd T_{00}\over \dd x^0}  h^{00}& =& {1\over 2}{\dd T^{00}\over \dd x^0}  h^{00}\nonumber \\
  & = & -{1\over 2}{\dd T^{0i}\over \dd x^i}  h^{00}
\rightarrow {1\over2} T^{0i} {\dd h^{00}\over \dd x^i}\nonumber\\
&  \simeq  & - {\rho}\bb{v}\bcdot  \del \Phi,
\end{eqnarray}
which is the rate at which the effective Newtonian potential $\Phi$ does {\it net} work on the matter.  (Here, $\rho$ is the Newtonian mass density and $\bb{v}$ the normal kinetic velocity.  Averaging is understood; the $\langle\rangle$ notation has been suppressed in [\ref{32}] for ease of presentation.)   This is negative
if the force is oppositely directed to the velocity, so that the source is losing energy by generating outgoing waves.   Our TT gauge expression (\ref{ttg}) for ${\cal T}_{0i}$ is also negative for an outward flowing wave of argument  $(r- t)$,
$r$ being spherical radius and $t$ time.   (By contrast, ${\cal T}^{0i}$ would be positive.)   

A subtle but important point:  can one be sure that such a potential actually exists?   An ordinary Newtonian potential would conserve mechanical energy over the course of the system's evolution.   That a gauge does exist in which an appropriate effective potential function emerges is shown in 
standard texts \cite{mis, mag}.   This is the Burke-Thorne potential\cite{bur, thor}, which is proportional to the leading order ``radiation reaction'' term in an expansion of $h^{00}$.  The effective potential must emerge as part of the radiation reaction terms in $\hmn$ if it is to deplete mechanical energy.    This time-dependent potential may be precisely defined in a suitable ``Newtonian gauge''.   Although we make no explicit use of it here, it is given by the following \cite{mis}:
$$
\Phi = -{1\over 2} h_{00} = {G\over 5} I^{(V)}_{jk} x_j x_k
$$
where $I_{jk}$ is the traceless moment of intertia tensor, $I^{(V)}$ refers to the fifth time derivative, and $x_j$ is a spatial
Cartesian coordinate. 
This justifies our normalization factor of $1/2\kk$.
Our final energy equation in an arbitrary gauge thus takes the form:
\beq\label{energy}
{\dd{\cal U}_{\rho\sigma}\over \dd x_\rho} = -{1\over 2}T_{\mu\nu} {\dd  h^{\mu\nu}\over \dd x^\sigma},
\eeq
The fact that the Newtonian gauge is not harmonic may have contributed to this rather basic (work done)$\leftrightarrow$(wave flux) conservation equation, the
analog of our scalar prototype introductory example, not being highlighted previously in the literature.   
If, for example, we follow custom and go directly to an harmonic gauge straight from equation (\ref{1}),
one obtains the following familiar result:
\beq
\Box \hmn=-\kk\Tmn\qquad {\rm (harmonic\ gauge)}.
\eeq
This is certainly useful as a means to solve for $\hmn$, but if we now multiply by $\dd_\sigma\bar h^{\mu\nu}$, and regroup as before,
we find the following: 
\beq
{\dd\ \over \dd x_\rho} \left( {1\over 2\kk}
{\dd\hmn\over  \dd x^\rho}   {\dd{\bar h}^{\mu\nu}\over \dd x^\sigma}\right)   = -{1\over 2}\left(\dd\bar h^{\mu\nu}\over\dd x^\sigma\right)\Tmn\qquad {\rm (harmonic\ gauge).}
\eeq
The difficulty now is that the the right side has no
obvious physical interpretation, and we are gauge-bound.   It is only if we follow the path
of equation (\ref{energy}), retaining full gauge freedom, that we may simultaneously formulate a conserved flux on one side
of the equation with the readily interpretable ``work done'' combination {\color {black} $-(T_{\mu\nu}/2) \dd_\sigma h^{\mu\nu}$} on the other.   Within the same equation, the radiated waves and the effective Newtonian potential are best understood
each in their own gauge.    Gauge selection (as an aid to interpretation) is the last, not the first, step of the process.  Very
different gauges in very different regions are illustrative of the nonlocal character of this problem.



\subsection{ Index symmetry and angular momentum conservation}
The tensor ${\cal U}_{\rho\sigma}$ lacks index symmetry for nonharmonic gauges.   This is awkward for angular momentum conservation.  In equation (\ref{energy}), set the index $\sigma = k$, a spatial index.   Next,
multiply the equation by $\eps_{ijk}x_j$, adhering to the usual summation convention for the
spatial indices but not distinguishing between their covariant and contravariant placement.   An integration
by parts then gives the following:
\beq
{\dd{\cal J'}_{i\rho}\over \dd x^\rho} -\eps_{ijk}{\cal U}_{jk}= -{1\over 2}\eps_{ijk} x_j T_{\mu\nu} {\dd  h^{\mu\nu}\over \dd x_k},
\eeq
where we have introduced a provisional angular momentum flux of gravitational waves
\beq
{\cal J'}_{i\rho} \equiv\eps_{ijk} x_j {\cal U}_{\rho k}.
\eeq
Were ${\cal U}_{\rho\sigma}$ a symmetric tensor, $\eps_{ijk}{\cal U}_{jk}$ would vanish identically,
and an equation of strict angular momentum would emerge.    This is an indication that the asymmetry 
is not a true physical asymmetry.  

Indeed, symmetry is easily
restored.   The method is to subtract off an appropriate ``difference tensor'' from ${\cal U}_{\rho\sigma}$ which leaves the fundamental conservation equations intact.    Begin
by rewriting ${\cal U}_{\rho\sigma}$ as
\beq\label{uu}
{\cal U}_{\rho\sigma} = t_{\rho\sigma}  + \eta_{\rho\sigma}{\cal S} +{1\over 2\kk} {\dd {\bar h}^{\lambda\mu}\over  \dd x^\lambda}\left( {\dd {\bar h}_{\mu\sigma}\over \dd x^\rho}-  {\dd {\bar h}_{\mu\rho}\over \dd x^\sigma}\right)\equiv
t_{\rho\sigma} + {\cal U}^D_{\rho\sigma},
\eeq
where $t_{\rho\sigma}$ is the standard symmetric wave tensor given by equation (\ref{tt}) and ${\cal U}^D_{\rho\sigma}$, the (spatially averaged) difference tensor, is defined by this equation.   It is easy to verify that ${\cal U}^D_{\rho\sigma}$ vanishes for the TT gauge (we have already done so), but it is also a gauge-invariant quantity under the infinitesimal coordinate transformation $x^\lambda\rightarrow x^\lambda + \xi^\lambda$, with the following:
\beq
\hmn\rightarrow\hmn -{\dd\xi_\mu\over \dd x^\nu}-{\dd\xi_\nu\over \dd x^\mu}+\eta_{\mu\nu} {\dd \xi^\lambda\over\dd x^\lambda}.
\eeq
Here $\xi_\mu$ is a well-behaved but otherwise arbitrary vector function.  In fact, it is a straightforward exercise to show that ${\cal S}$ and the final $\rho\sigma$-antisymmetric term in (\ref{uu}) are each gauge invariant on their own.   A standard textbook problem to show that $t_{\rho\sigma}$ is gauge-invariant\cite{mis}; here we have done so indirectly, because ${\cal U}_{\rho\sigma}$ must be gauge invariant by virtue of its original construction.    Thus,
if we evaluate ${\cal U}^D_{\rho\sigma}$ in the TT gauge, in which it vanishes, and transform to any other gauge, the result must still be zero.   We may thus conclude that ${\cal U}_{\rho\sigma}=t_{\rho\sigma}$ quite generally.   

Returning to the question of angular momentum conservation, we replace ${\cal U}_{\rho\sigma}$
with $t_{\rho\sigma}$, and define the symmetrized angular momentum flux tensor as follows:
\beq
   {\cal J}_{i\rho} \equiv\eps_{ijk} x_j {t}_{\rho k}.
\eeq
The precise statement of angular momentum conservation is then as follows:
\beq
{\dd{\cal J}_{i\rho}\over \dd x^\rho} = -{1\over 2}\eps_{ijk} x_j T_{\mu\nu} {\dd  h^{\mu\nu}\over \dd x_k}.
\eeq
The right side affords a direct method for computing angular momentum loss via the explicit Burke-Thorne potential.

\section {Conclusion}
The linear wave equation that emerges from the Einstein field equations, either in the form of (\ref {1}) or (\ref{eq}), contains in itself all the ingredients needed for determining a conserved gravitational wave energy flux tensor, propagating in a background Minkowski spacetime and produced by slowly moving sources.   The stress tensor that is calculated via a more lengthy and complex second-order analysis of the Einstein tensor 
is, for any harmonic gauge, identical to that which energes from our first order calculation, i.e. 
${\cal U}_{\rho\sigma}$ and $t_{\rho\sigma}$ are identical in this case.    It is only for the construction of a symmetric wave stress
tensor in nonharmonic gauges that an alteration of form is needed.    

The presented calculation also illuminates the physical connection between the radiated gravitational waves and the effective Newtonian potential that serves to deplete mechanical energy from the matter source for these waves.    
The precise form of this
``Burke-Thorne'' potential does not itself play a role in our analysis.  Merely the fact that it exisits, and that in common with any Newtonian potential function it is associated with $-h^{00}/2$ to
leading order, is sufficient to determine the normalization constant of the conserved flux tensor.    Indeed, the entire calculation
could be performed in an harmonic gauge, in which case $-(T_{\mu\nu}/2)\dd_\sigma h^{\mu\nu}$ must be the generic expression for the work done on the sources, even if its manifestation is less transparent than in the Newtonian gauge.  

We have shown how to remove an apparent index asymmetry in ${\cal U}_{\rho\sigma}$,
which in its symmetrized and spatially averged form reverts to $t_{\rho\sigma}$.  Angular
momentum conservation readily follows.

The approach that is presented in this paper seems to be the simplest, the most concise, and ultimately the most physically transparent route to understanding the form of the energy flux of gravitational radiation, especially in its most natural harmonic gauges, as embodied in equation (\ref{summ}).

\section*{Acknowedgements}

I am most grateful to J.\ Binney and P.\ Ferreira for a critical readings of an early draft of this work, and for their many 
helpful suggestions.      It is likewise a pleasure to acknowledge stimulating conversations with R. Blandford, P. Dellar, C. Gammie, M. Hobson, D. Lynden-Bell, J. Magorrian, C. McKee, J. Papaloizou and W. Potter.   Finally, it is a pleasure to thank the referees T.\ Baumgarte and J.-P.\ Lasota for their excellent advice and support.  I acknowledge support from a gift from the Hintze Charitable Fund, from the Royal Society in the form of a Wolfson Research Merit award, and from the Science and Technology Facilities Council.


\begin{thebibliography}{99}
\bibitem{note} 
In this paper, Greek indices indicate spatio-temporal dimensions; Roman
indices spatial dimensions.  We use the sign convention $[-+++]$ for the diagonal Minkowski metric $\eta_{\mu\nu}$. Indices of
$h_{\mu\nu}$ [$h^{\mu\nu}$] are raised [lowered] with $\eta^{\mu\nu} [=\eta_{\mu\nu}]$.   For in-line equations, we use
the notation $\dd^\mu \equiv \dd/\dd x_\mu$, $\dd_\mu\equiv \dd /\dd x^\mu$.
\end{thebibliography}
\end{document}